\documentclass[aps,prd,amsmath,amsfonts,a4paper,11pt,reprint,twocolumn,square,numbers,showpacs,superscriptaddress,floatfix,sort&compress]{revtex4-1}
\usepackage{hyperref}
\usepackage{amssymb}
\usepackage{bm}
\usepackage{natbib}
\usepackage{graphicx}
\expandafter\let\csname equation*\endcsname\relax
\expandafter\let\csname endequation*\endcsname\relax
\usepackage{amsmath}
\usepackage[labelfont={bf,scriptsize},textfont={scriptsize},justification=RaggedRight,format=hang]{caption,subfig}
\usepackage{color}

\RequirePackage{ifpdf}
\ifpdf
\else 
\fi

\usepackage{titlesec}
\titlespacing*{\section}{0pt}{0.6cm}{0.5cm}

\newcommand{\para}[1]{\par\vspace{2mm}\noindent\textbf{\emph{{#1}}---}}

\renewcommand{\d}{\mathrm{d}}

\newcommand{\vect}[1]{\bm{\mathrm{{#1}}}}

\newcommand{\Mp}{M_{\mathrm{Pl}}}

\newcommand\be{\begin{equation}}
\newcommand\ee{\end{equation}}
\newcommand{\bea}{\begin{eqnarray}}
\newcommand{\eea}{\end{eqnarray}}

\begin{document}

\begin{center}
\rightline{\small DESY-16-073}
\vskip -3cm
\end{center}

	\title{Simple emergent power spectra from complex inflationary physics}

	\author{Mafalda Dias}
	\email{mafalda.dias@desy.de}
	\affiliation{Deutsches Elektronen-Synchrotron, DESY, Notkestra\ss e 85, 22607 Hamburg, Germany}

	\author{Jonathan Frazer}
	\email{jonathan.frazer@desy.de}
	\affiliation{Deutsches Elektronen-Synchrotron, DESY, Notkestra\ss e 85, 22607 Hamburg, Germany}

	\author{M.C.~David Marsh}
	\email{m.c.d.marsh@damtp.cam.ac.uk}
	\affiliation{Department of Applied Mathematics and Theoretical Physics, DAMTP, University of Cambridge,
Cambridge, CB3 0WA, United Kingdom}
		
	\begin{abstract}
We construct ensembles of random scalar potentials for $N_f$ interacting scalar fields using non-equilibrium random matrix theory, and use these to study the generation of observables during small-field inflation. 
For $N_f={\cal O}({\rm few})$, these heavily featured scalar potentials  give rise to power spectra that are highly non-linear, at odds with observations.
For $N_f\gg 1$, the superhorizon evolution of the perturbations is generically substantial, yet the power spectra simplify considerably and become more predictive, with most realisations being well approximated by a linear power spectrum. This provides proof of principle that complex inflationary physics can give rise to simple emergent power spectra. 
We explain how these 
results can be understood in terms of 
large $N_f$ universality of random matrix theory.
	\end{abstract}
	
	\maketitle
		According to the cosmological paradigm of inflation~\cite{Guth:1980zm,Linde:1981mu,Albrecht:1982wi},  the structure of the observed universe emerged from the gravitational collapse of small, primordial density perturbations, which in turn originated as quantum fluctuations during a period of accelerated expansion. The cosmological parameters inferred from observations of the Cosmic Microwave Background radiation (CMB) are in striking agreement with the predictions of many of the simplest models of single-field slow-roll inflation \cite{Planck:2013jfk}: at the present accuracy of cosmological surveys, the spectrum of adiabatic curvature perturbations is Gaussian and almost scale-invariant and can thus be described with just two numbers --- the amplitude of the power spectrum $A_{s}$ and its tilt $n_{s}$. Should then the simplicity of the observed CMB perturbations be interpreted as evidence for a minimal and simple microscopic origin of inflation in fundamental theory?
	
Attempts at embedding inflation in string theory have revealed that even ostensibly simple inflationary models require  delicate 
arrangements of the various sources of the scalar potential, which in general is a complicated function of a large number of scalar `moduli' fields  \cite{Baumann:2014nda}. More general inflationary models may 
involve many dynamically important fields  with complicated interactions, however the explicit construction of such  
models is very challenging, and not much is known about the observational predictions of complex many-field models of inflation.

In this paper we address this question, and provide proof of principle that a complex many-field inflationary model can give rise to simple power spectra. By computing the power spectra of inflationary models with highly featured, randomly generated scalar potentials  of between 2 and 50 scalar fields, we find that systems with a large number of fields generate simpler and far less featured power spectra during inflation. We interpret this result using Random Matrix Theory (RMT),  finding that in systems with many fields, eigenvalue repulsion makes the power spectra both more predictive and better approximated by a linear power-law. RMT universality  then suggests that these results should be   
applicable well beyond the class of random inflationary models that we study explicitly~\footnote{There have been attempts at finding simple predictions in models with a few scalar fields, by identifying single field attractor solutions \cite{Frazer:2013zoa, Kaiser:2013sna, Kallosh:2013daa}. In the context of N-flation, it has also been shown that predictions can simplify in the large $N_{f}$ regime~\cite{Easther:2013rva,Bachlechner:2014hsa}. In contrast to the setup studied in this paper however, the N-flation potential has an exceedingly simple functional form.}.  
	
	\section*{THE MODEL}	

To study the impact of complexity on observables, we 
construct ensembles of 
 randomly generated scalar potentials suitable for inflation  using
the method presented in~\cite{DBM}. 
By constructing each scalar potential locally around the  
field trajectory, the computational cost (which for complicated, globally constructed potentials quickly becomes prohibitive \cite{Frazer:2011br, Easther:2013rva})
 is minimised, 
  thus enabling us to study inflation in potentials with structure on scales $\ll \Mp$ for a large number of fields.
We will consider  statistically  isotropic systems, and stipulate that the collection of Hessian matrices associated with a set of well-separated points in field space constitute a random sample of the Gaussian Orthogonal Ensemble (GOE). According to \cite{DBM} (see also \cite{Battefeld:2014qoa}), such potentials can be constructed using the non-equilibrium RMT technique of Dyson Brownian Motion (DBM) \cite{1962JMP.....3.1191D}. 

More precisely, for a field located at 
a point $p_0$ in field space, we write the potential locally as a Taylor expansion to quadratic order 
by specifying the value of the potential  $V|_{p_0}$, its gradient $\partial_a V|_{p_0}$, and the Hessian matrix $\partial^2_{ab} V|_{p_0}$.  
We choose these initial conditions to be favourable for inflation, in particular, we focus on the rare regions in which the potential is locally very flat, and we set the initial field space  velocity $\dot{\phi}^a|_{p_0}$ to be suitable for slow-roll inflation. Thus, close to $p_0$ the potential is well-approximated by,
\be
V\Big|_{p_0} = \Lambda_{\rm v}^4 \sqrt{N_f} \left(
v_0 |_{p_0}+ v_a|_{p_0}\, \tilde{\phi}^a + \frac{1}{2} v_{ab} |_{p_0}\, \tilde{\phi}^a\tilde{\phi}^b
\right),
\label{eq:V}
\ee
where $ \Lambda_{\rm v}$ defines the `vertical'  scale of the potential, and $\tilde{\phi}^a \equiv \phi^a / \Lambda_{\rm h}$ are  dimensionless fields with  $ \Lambda_{\rm h}$ setting the scale over which the potential exhibits structure. The overall factor of $\sqrt{N_f}$ is explained in \cite{DBM}.

We then numerically 
solve the dynamical equations of motion of this system over a path length $\delta s\ll  \Lambda_{\rm h}$ to the point $p_1$, at which point 
we update the coefficients of the Taylor expansion to find the new local description of the potential. 
To linear order in Taylor expansion we then find the scalar potential and gradient at $p_1$ to be given by $v_0|_{p_1} = v_0|_{p_0} + v_a|_{p_0} \delta s^a/\Lambda_h$, and $v_a|_{p_1} = v_a|_{p_0} + v_{ab}|_{p_0} \delta s^b/\Lambda_h$, while, 
\be
v_{ab}|_{p_1} =v_{ab}|_{p_0} + \delta v_{ab}|_{p_0\to p_1} \, ,
\ee
where $\delta v_{ab}$ is a small, stochastic matrix perturbation obtained by `Dyson Brownian motion' \cite{1962JMP.....3.1191D}. The first two moments of $\delta v_{ab}$ are given by,
\bea
\langle \delta v_{ab}|_{p_0\to p_1}\rangle &=& - v_{ab}|_{p_0}  \frac{\delta s}{ \Lambda_{\rm h}} \, , \\
\langle (\delta v_{ab}|_{p_0\to p_1})^2 \rangle &=& (1+\delta_{ab}) \frac{\delta s}{ \Lambda_{\rm h}} \sigma^2 \, , 
\eea
where $\delta s = \sqrt{\delta s_a \delta s^a}$ and $\sigma^2$ sets the variance of the distribution (we  take $\sigma^2=2/N_f$ so that for $N_f\gg1$, the spectrum of $v_{ab}$ is $N_f$ independent).
Having found the  values of the scalar potential, gradient, Hessian and field space velocity at $p_1$, we may re-iterate the procedure 
and evolve the system to another nearby point $p_2$. This way we stitch together the random scalar potential patch by patch along the dynamically determined field trajectory. As the field evolves to a distance $s$ away from the initial point $p_0$, due to eigenvalue repulsion (cf.~Fig. 3 of \cite{DBM}) the Hessian matrix relaxes from a rare fluctuated spectrum, suitable for inflation,
towards the Wigner distribution which is the typical spectrum of the GOE~\cite{1962JMP.....3.1191D, Uhlenbeck:1930zz}. We terminate the iterative process once inflation ends, defined to be when the slow-roll parameter $\epsilon \equiv -\dot{H}/H^2=1$. This construction is applicable to both small-field $s < \Mp$ and large-field $s > \Mp$ inflation but in this paper we will only study the $s < \Mp$ regime. The construction does not require an underlying shift symmetry and hence the method seems particularly well suited to the study of small-field inflation.


	\section*{METHOD}

	\para{Computing $P_{\zeta}(k)$ } Computing perturbations for large $N_f$ models is in general a numerically heavy task, as it involves an integration from deep inside the horizon up until the end of inflation. The main computational expense comes from the fact that in order to obtain the two-point function 
		of the curvature perturbation, $\zeta$, one is required to solve $\mathcal{O}(N_f^2)$ coupled ordinary differential equations \footnote{These can be the equations of motion either for field-field, field-momenta and momenta-momenta correlations function, as in \cite{2pf, optics, mulryne}, or for the two-index mode functions \cite{Salopek, Ringeval, Huston, modecode}.}.
	Furthermore, this system of equations has an explicit $k$-dependence, meaning that the full power spectrum of $\zeta$, $P_\zeta(k)$, can only be obtained by calculating the amplitude for each mode individually.

Here, we use the patchwork construction of the DBM potential to find striking simplifications  for slow-roll multifield inflation, enabling us to bypass these problems. 
	The key step is to rotate the field basis independently for each patch to the local eigenmodes, $\varphi^a$, of the Hessian. The great advantage in doing so is that the potential in each patch is then (locally) sum-separable $V(\varphi^{1},\ldots,\varphi^{N_f})=\sum_{a}^{N_{f}}V_a(\varphi^{a})$.
In this case, the field perturbations in a spatially flat gauge evolve like,
\be
\delta \varphi^a \big|_{p_{i+1}} = \Gamma^a_{~b} (p_{i+1},p_i) \delta\varphi^b |_{p_i} \, ,
\ee
where, assuming the slow-roll equations of motion are valid, the propagator $\Gamma^a_{~b} (p_{i+1},p_i)$ can be expressed purely in terms of background quantities \footnote{We will present the derivation in a separate paper \cite{FutureWork} but a similar calculation can be found in Refs.~\cite{vernizzi,battelfeld}.},  thus providing an analytic solution to the propagation of the perturbation over each patch.

The full evolution of the field perturbations until the end of inflation at the point $p_f$ is then simply given by 
the path-ordered product of propagators and orthogonal transformations,
\bea
\delta \vec{\phi} \big|_{p_f} &=& 
O^{\rm T}_{p_{f}}\Gamma(p_{f},p_{f-1}) O_{p_{f}}\, 
\ldots O^{\rm T}_{p_{1}}\Gamma(p_{1},p_{0}) O_{p_{1}}\,  \delta \vec{\phi} \big|_{p_{0}} \nonumber \\
&\equiv&
\Gamma(p_f, p_0) \, 
\delta \vec{\phi} \big|_{p_0} 
 \, .
\eea
The curvature perturbation at the constant density surface,
$\zeta$,   is then obtained by a standard gauge transformation in the final patch,
\be
\zeta = N_{a}  \delta \phi^{a} |_{p_f} \, .
\ee 
where $N_{a}\equiv \partial N / \partial\phi^{a}|_{p_f}$. It follows 
that the two point function,
\begin{equation}
    \label{eq:generic-2pf}
    \langle \delta\phi^a(\vect{k}_1) \delta\phi^b(\vect{k}_2) \rangle
    =
    (2\pi)^3 \delta(\vect{k}_1 + \vect{k}_2) \frac{\Sigma^{ab}}{k^3} \, ,
\end{equation}
evolves according to two copies of the propagator \cite{optics}. Hence the power spectrum may be expressed as,
\begin{equation}
    P_{\zeta}(N)=N_{a}N_{b}{\Gamma^{a}}_{c}{\Gamma^{b}}_{d}\Sigma^{cd}(N_{*}) \, ,
\end{equation}
where we assume the field perturbations crossing the horizon at time $N_{*}$ are uncorrelated, 
\be\label{eq:SigmaHcrossing}
\Sigma^{ab}(N_{*})=\frac{H^{2}_*}{2}\delta^{ab} \, ,
\ee
an approximation we find to be in excellent agreement with our numerical tests.
With this method, the two-point correlation function can be computed purely from information already obtained in the process of constructing the potential and it is trivial to obtain the full spectrum $P_\zeta(k)$. This method will be described in full detail in \cite{FutureWork}. 

The spectral index $n_{s}$ and its running $\alpha_{s}$,
\begin{equation}
n_{s}-1\equiv \frac{\d \ln P_{\zeta}}{\d \ln k} \, ,\quad \quad \alpha_{s}\equiv \frac{\d n_{s}}{\d \ln k} \, ,
\end{equation}
can likewise be expressed in terms of the propagator $\Gamma^a_{~b}$.
Combining results from \cite{optics, transportns}, the spectral index is given by,
\begin{align}\label{eq:ns}
n_s - 1 &= \frac{1}{P_{\zeta}} N_a N_b {\Gamma^a}_c {\Gamma^{b}}_d n^{cd}_*
\, ,
\end{align}
where, 
\begin{align}
n^{ab}_* &\equiv \frac{\d \Sigma}{\d \ln k}\bigg|_{*} = (-\epsilon\delta^{ab}-u^{ab})_* H_*^2 \, ,
\end{align}
and the matrix $u_{ab}= \partial^2_{ab} \ln V$. Similarly, the running takes the form \cite{dbrane},
\begin{equation}\label{eq:alpha}
\alpha_{s}=\frac{1}{P_{\zeta}} N_a N_b {\Gamma^a}_c{\Gamma^{b}}_d \alpha^{cd}_*-(n_{s}-1)^{2} \, ,
\end{equation}
where, 
\begin{align}
\alpha^{ab}_* \equiv& \frac{\d n^{ab}}{\d \ln k}\bigg|_{*}\\
=& [(2 \epsilon^{2}-\epsilon')\delta^{ab}- u'^{ab}+2\epsilon u^{ab}]_*H_*^2-2[u^{a}_{~c} n^{cb}]_*  \nonumber \, ,
\end{align}
and primes indicate differentiation with respect to e-fold time $N$.

	\para{Numerical procedure } 
	We are now ready to compute the spectrum of the curvature perturbation generated during inflation in random multifield models constructed through Dyson Brownian Motion. The potential~\eqref{eq:V} has a number of parameters: the number of fields $N_f$, the vertical and horizontal scales, $ \Lambda_{\rm v}$ and $ \Lambda_{\rm h}$, and the initial conditions for the potential $v_0|_{p_0}$, $v_a|_{p_0}$ and $v_{ab}|_{p_0}$. In this work we focus solely on effects emerging from large $N_f$ behaviour, leaving a more exhaustive study of the full parameter space to future work \cite{FutureWork}.

	For the remaining parameters we take  $v_0|_{p_0}=1$, and chose $v_a|_{p_0}$  to set the initial value of the $\epsilon$ slow-roll parameter. The initial spectrum of $v_{ab}|_{p_0}$ is chosen to be that of a fluctuated Wigner spectrum \cite{Dean:2006wk} with an approximately vanishing smallest eigenvalue, taking the eigenvector of the smallest eigenvalue of $v_{ab}|_{p_0}$ to be aligned with $v_a|_{p_0}$. We note that  eigenvalue repulsion quickly modifies the initial spectrum during Dyson Brownian Motion, leading to mass spectra with features on scales $\ll  \Lambda_{\rm h}$, and an insensitivity to the details of the  initial distribution of $v_{ab}$ \cite{DBM}. The models we consider are then of small-field  `approximate saddle-point'-type with $ \Lambda_{\rm h}<\Mp$.
	
	Finally, 
	for random potentials that give rise to at least 60 e-folds of inflation, we compute $P_\zeta(k)$ for the scales leaving the horizon between 50 and 60 e-folds before the end of inflation; assuming it is approximately this 10 e-fold range which is constrained by observations of the CMB. The `vertical scale', $ \Lambda_{\rm v}$, is chosen to set the  amplitude of the power spectrum of the mode $k_0$ exiting the horizon 55 e-folds before the end of inflation to agree with the COBE normalisation \footnote{Note that the slow-roll equations of motion are independent of $ \Lambda_{\rm v}$, thus $ \Lambda_{\rm v}$ can be adjusted to normalise the power spectrum.}. 
	
	\section*{RESULT}	
	
\begin{figure}[b]
  \includegraphics[width=0.43\textwidth]{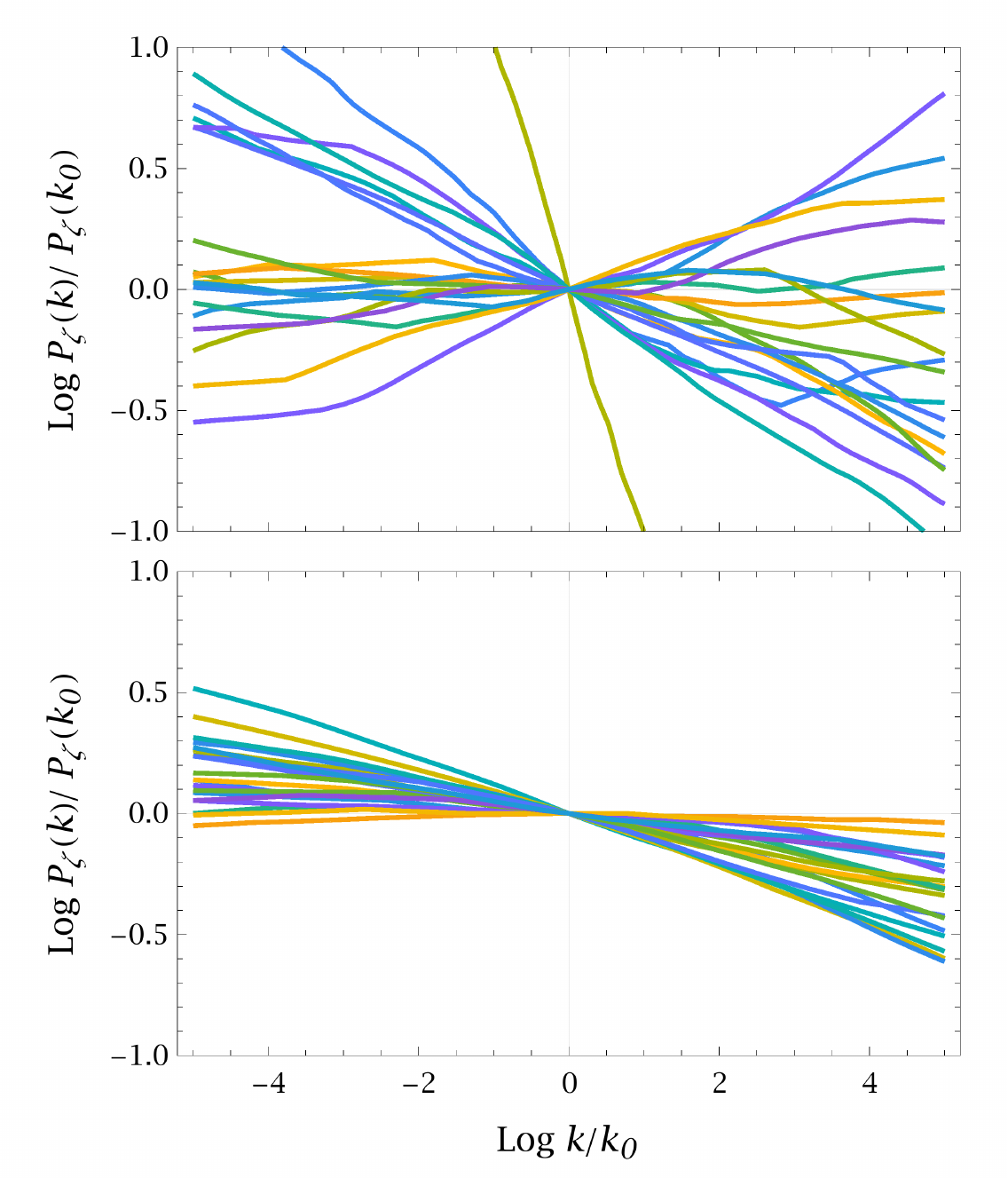}
  \caption{Example power spectra for the scales leaving the horizon between 50 and 60 e-folds before the end of inflation for $N_f=2$ (top) and $N_f=50$ (bottom), with $ \Lambda_{\rm h}=0.4$. }
   \label{fig:Pzetaofkplots}
\end{figure}
	
Fig.~\ref{fig:Pzetaofkplots} summarises our main result. Displayed are the power spectra of a random selection of 25 inflationary realisations for $N_f=2$ and $N_f=50$, with all other parameters fixed.  When the number of fields is small (top), the power spectra vary dramatically between realisations and are typically highly non-linear \footnote{Here and throughout we are concerned with whether or not $\log P_{\zeta}(\log k)$ is linear.}. In contrast, when many fields are active during inflation (bottom), the spectra become much \emph{simpler} and can generically  
 be well described by a linear fit. 
Moreover, at large $N_f$ the distribution of the spectra also becomes less varied and more \emph{predictive}, with sharper distributions for the spectral index and its running.  In fact, for sufficiently large $ \Lambda_{\rm h}$,  
the spectra generated during inflation become consistently too  red to match observational constraints. 

\begin{figure}
  \includegraphics[width=0.43\textwidth]{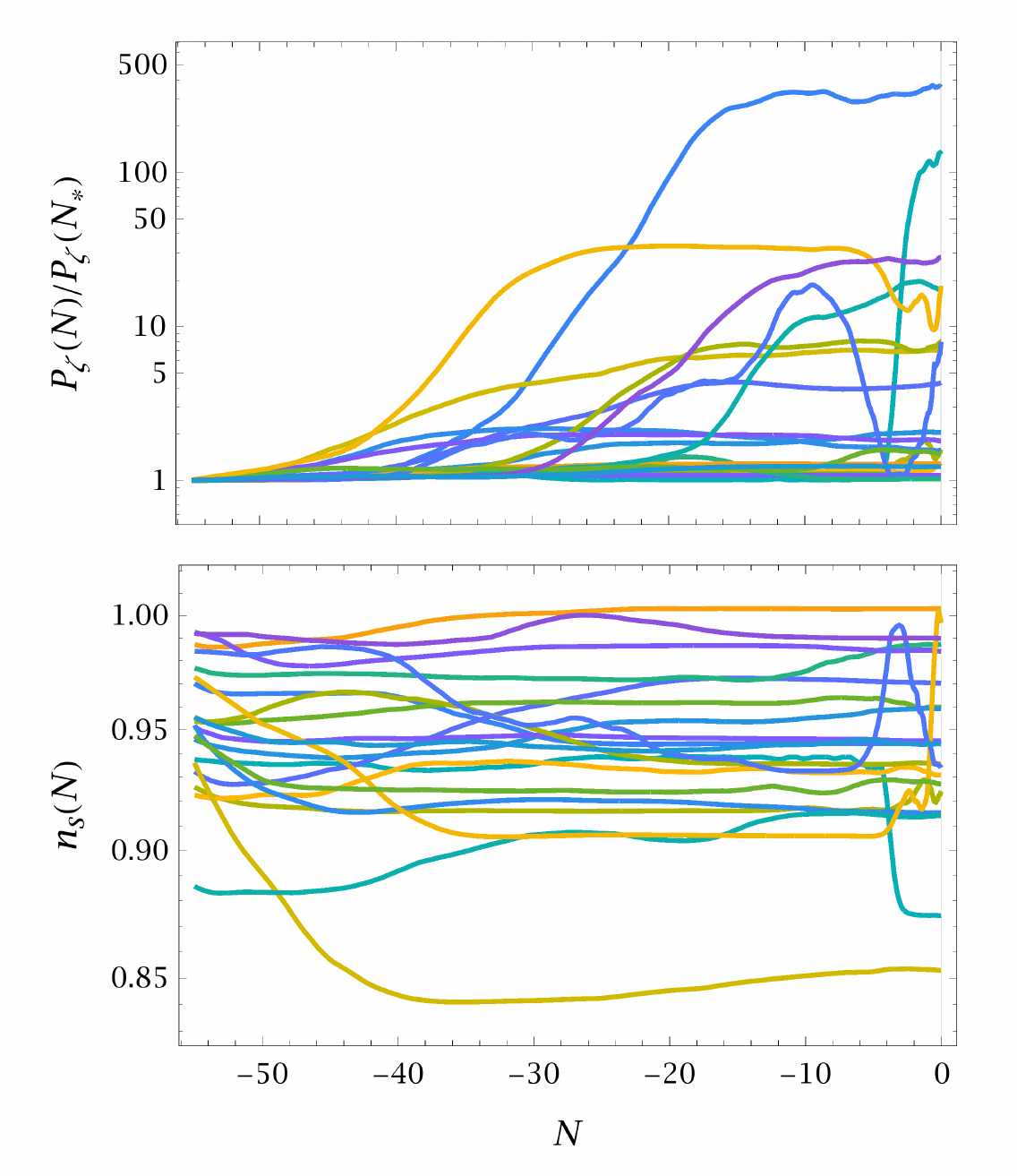}
  \caption{Superhorizon evolution of $P_\zeta$ (top) and $n_{s}$ (bottom) for the $k_0$ mode, which leaves the horizon 55 e-folds before the end of inflation, for the examples with  $N_f=50$ of Fig.~\ref{fig:Pzetaofkplots}. The power spectra are normalised to their horizon exit value. $N=-55$ corresponds to the horizon exit time and $N=0$ to the end of inflation.}
     \label{fig:zetaofNplots}
\end{figure}

The emergent simplicity at large $N_f$ does not however imply that this limit corresponds to an effectively single field regime.
Fig.~\ref{fig:zetaofNplots} shows the evolution of $P_{\zeta}(k_0)$ and $n_{s}(k_0)$ on superhorizon scales for the same $N_f = 50$ realisations shown in Fig.~\ref{fig:Pzetaofkplots}. 
We find the amount of superhorizon evolution to always be substantial, often changing the amplitude of the spectrum by several orders of magnitude. As superhorizon evolution of $\zeta$ occurs due to the transfer of power from isocurvature to adiabatic perturbations, this is a direct indication of multifield dynamics. Conversely, for $N_f=2$ the 
evolution of $P_{\zeta}(k_0)$ on superhorizon scales is, up to numerical accuracy, zero (hence no plots are shown).

All inflationary realisations considered here are of small-field type with typical path length $\|\Delta\phi\| \sim 0.7 \Lambda_{\rm h}$, immediately implying a small value for the tensor-to-scalar ratio, $r$, according to the Lyth bound, $r \lesssim 0.01 \left( \|\Delta \phi\|/\Mp \right)^2$ \cite{Lyth:1996im}. Moreover, due to the sharp increase of $\epsilon$ during this type of approximate saddle-point inflation and, for large $N_f$, the superhorizon transfer of power into the scalar perturbations, this bound   is far from saturated: for $N_f=2$, $r\approx 10^{-7}$ and for $N_f=50$, $r\approx 10^{-10}$. The respective typical values of $H_{*}$ are $\sim10^{11}$ GeV and $\sim 10^{9}$ GeV.

\section*{DISCUSSION} 
Strikingly, our simulations of random and complicated inflationary models show that at large $N_f$ the ensemble of inflationary realisations become more predictive, and the generated power spectra become approximately linear. 
We now explain how  these effects can be understood in terms of random matrix theory. 

\para{Improved predictability at large $N_{f}$ } 
The variance of the spectral index for an ensemble of random potentials is a
 measure of the predictability of the model.
As mentioned, all our inflationary realisations are of small-field type and have very small $\epsilon$ at horizon crossing. In particular, terms in Eq.~\eqref{eq:ns} proportional to $\epsilon_*$ and gradients $\partial_a V_*$ are subdominant and the overall expression
simplifies considerably,
\begin{equation}
n_s - 1 \approx 2 e_a e_b \left (\frac{v^{ab}}{v_{0}\Lambda_{\rm h}^{2}} \right )_{*} \, .
\end{equation}
Here, the unit vector  $e_{a}\equiv N_{b}\Gamma^{b}_{a}/\| N_{c}\Gamma^{c}_{d}\|$ evolves throughout inflation, but tends to primarily develop non-negligible 
 components in the directions of the first few smallest eigenvalues of $v_{ab}|_{*}$, which then dominate the contributions to $n_s$.
The variances of the smallest few eigenvalues of $v_{ab}|_{*}$, and in particular that of the smallest eigenvalue $\lambda_1$, are then the main contributions to the variance of the spectral index. 
Sufficiently close to the initial patch, we can estimate the first two moments of $\lambda_1$ as follows: at the $k$:th patch, $v_{ab}^k = v_{ab}^0 + \sum_{l=1}^{k} \delta v_{ab}^l$, so that 
to second order perturbation theory,
\be
\lambda_1^k = v_{11}^k - \sum_{b'=2}^{N_f} c_{b'} |v_{1b'}^k|^2  \, , \label{eq:2opt}
\ee
where $c_{b'} = |\lambda_1^0 - \lambda_{b'}^0|^{-1}$.
Computing the first two moments of $\lambda^k_1$ and then taking the  continuum limit ($k\to\infty$, $s$ fixed), we find, 
$
\langle \lambda_1(s) \rangle = e^{-s} \lambda_1^0 -s \sigma^2 \sum_{b'=2}^{N_f} c_{b'} 
$, and,
\be
{\rm Var}\left(\lambda_1(s)\right) = 2 s \sigma^2 \left(1+ s \sigma^2  \sum_{b'=2}^{N_f} c_{b'}^2\right) \, . \label{eq:var}
\ee
Thus, eigenvalue repulsion drives the mean of the smallest eigenvalue to negative values,  
explaining the preference for red spectra. 
For fixed, small $s$, the overall prefactor of \eqref{eq:var} decreases as $ \sigma^2 \sim 1/N_f$, indicating a shrinking variance with increasing $N_f$, and hence an increased predictivity of the model, in agreement with numerical simulations. The second term of \eqref{eq:var} grows with $N_f$ and hence indicates that the  second order perturbation theory quickly becomes insufficient for very large systems: the actual  probability distribution of the smallest eigenvalue becomes sharper at large $N_f$ due to strong eigenvalue repulsion.

\para{Smoother spectra at large $N_{f}$ }
Eigenvalue repulsion also explains the smoothening of the power spectra at large $N_f$: 
when $N_f$ is small, the smallest eigenvalue undergoes Brownian motion with a high volatility, while for large $N_f$ large fluctuations become increasingly rare, in agreement with our discussion of Eq.~\eqref{eq:var}.  
Hence, 
  the running $\alpha_{s}$, which is a measure of deviation from linearity, decreases with $N_{f}$, as we now show. Since our realisations are always of the small-field type,  Eq.~\eqref{eq:alpha} is well approximated by,
\begin{equation}\label{eq:newalpha}
\alpha_s  \approx 4e_a e_b\frac{v^{a}_{~c} v^{cb}}{v^{2}_{0}\Lambda_{\rm h}^{4}}\bigg|_{*} - 4 \left ( \frac{e_a e_b v^{ab}_*}{v_{0*}\Lambda_{\rm h}^{2}} \right )^2 + 2e_a e_b\frac{{v^{ab}}'}{v_{0}\Lambda_{\rm h}^{2}}\bigg|_{*} \, .
\end{equation}
Following the same line of argument as above, for a qualitative estimate of the behaviour of $\alpha_s$, we assume $\vec{e}$ to be roughly aligned with the most tachyonic direction. In that case the first two terms of Eq.~\eqref{eq:newalpha} approximately cancel and the last term, which is effectively the rate of change of the smallest eigenvalue of $v_{ab}$, dominates. Hence we attribute the radical decrease of $\alpha_s$ to the decrease in volatility of the smallest eigenvalue at large $N_f$.
For large $N_f$ the running falls within the range allowed by current Planck data.

\para{Final remarks }
We have for the first time shown that small-field inflationary models with many fields coupled through random, highly featured potentials 
are capable of generating simple power spectra that can be compatible with observations. Large $N_f$  random matrix theory provides an intuitive explanation for the observed effects: both the smoothness and the enhanced predictivity of the spectra are simple consequence of eigenvalue repulsion, which becomes strong as $N_f$ grows large. A more detailed treatment of observational signatures, including a study of the superhorizon evolution of the 
 isocurvature modes and the possible signals in 
 the bispectrum and trispectrum remain open questions that we intend to address in future work.

\section*{ACKNOWLEDGMENTS}	                                             
	We thank Thomas Bachlechner,  Daniel Baumann, David Ciupke, Ben Freivogel, Alexander Westphal and Lucila Zarate for useful discussions. We particularly thank Liam McAllister for numerous valuable discussions and his involvement in the early stages of this project. M.D. is supported by the German Science Foundation (DFG) within the Collaborative Research Centre 676 ÓParticles, Strings and the Early UniverseÓ and by the ERC Consolidator Grant STRINGFLATION under the HORIZON 2020 contract no. 647995. JF is supported by the ERC Consolidator Grant STRINGFLATION under the HORIZON 2020 contract no. 647995 and a grant from the Simons Foundation. DM acknowledges support from a Starting Grant of the European Research Council (ERC STG Grant 279617). This work was supported in part by National Science Foundation Grant No. PHYS-1066293, and we are grateful for the hospitality of the Aspen Center for Physics and the Cornell Particle Theory group.

\bibliography{refs}

\begin{thebibliography}{34}%
\makeatletter
\providecommand \@ifxundefined [1]{%
 \@ifx{#1\undefined}
}%
\providecommand \@ifnum [1]{%
 \ifnum #1\expandafter \@firstoftwo
 \else \expandafter \@secondoftwo
 \fi
}%
\providecommand \@ifx [1]{%
 \ifx #1\expandafter \@firstoftwo
 \else \expandafter \@secondoftwo
 \fi
}%
\providecommand \natexlab [1]{#1}%
\providecommand \enquote  [1]{``#1''}%
\providecommand \bibnamefont  [1]{#1}%
\providecommand \bibfnamefont [1]{#1}%
\providecommand \citenamefont [1]{#1}%
\providecommand \href@noop [0]{\@secondoftwo}%
\providecommand \href [0]{\begingroup \@sanitize@url \@href}%
\providecommand \@href[1]{\@@startlink{#1}\@@href}%
\providecommand \@@href[1]{\endgroup#1\@@endlink}%
\providecommand \@sanitize@url [0]{\catcode `\\12\catcode `\$12\catcode
  `\&12\catcode `\#12\catcode `\^12\catcode `\_12\catcode `\%12\relax}%
\providecommand \@@startlink[1]{}%
\providecommand \@@endlink[0]{}%
\providecommand \url  [0]{\begingroup\@sanitize@url \@url }%
\providecommand \@url [1]{\endgroup\@href {#1}{\urlprefix }}%
\providecommand \urlprefix  [0]{URL }%
\providecommand \Eprint [0]{\href }%
\providecommand \doibase [0]{http://dx.doi.org/}%
\providecommand \selectlanguage [0]{\@gobble}%
\providecommand \bibinfo  [0]{\@secondoftwo}%
\providecommand \bibfield  [0]{\@secondoftwo}%
\providecommand \translation [1]{[#1]}%
\providecommand \BibitemOpen [0]{}%
\providecommand \bibitemStop [0]{}%
\providecommand \bibitemNoStop [0]{.\EOS\space}%
\providecommand \EOS [0]{\spacefactor3000\relax}%
\providecommand \BibitemShut  [1]{\csname bibitem#1\endcsname}%
\let\auto@bib@innerbib\@empty
\bibitem [{\citenamefont {Guth}(1981)}]{Guth:1980zm}%
  \BibitemOpen
  \bibfield  {author} {\bibinfo {author} {\bibfnamefont {A.~H.}\ \bibnamefont
  {Guth}},\ }\href {\doibase 10.1103/PhysRevD.23.347} {\bibfield  {journal}
  {\bibinfo  {journal} {Phys.Rev.}\ }\textbf {\bibinfo {volume} {D23}},\
  \bibinfo {pages} {347} (\bibinfo {year} {1981})}\BibitemShut {NoStop}%
\bibitem [{\citenamefont {Linde}(1982)}]{Linde:1981mu}%
  \BibitemOpen
  \bibfield  {author} {\bibinfo {author} {\bibfnamefont {A.~D.}\ \bibnamefont
  {Linde}},\ }\href {\doibase 10.1016/0370-2693(82)91219-9} {\bibfield
  {journal} {\bibinfo  {journal} {Phys.Lett.}\ }\textbf {\bibinfo {volume}
  {B108}},\ \bibinfo {pages} {389} (\bibinfo {year} {1982})}\BibitemShut
  {NoStop}%
\bibitem [{\citenamefont {Albrecht}\ and\ \citenamefont
  {Steinhardt}(1982)}]{Albrecht:1982wi}%
  \BibitemOpen
  \bibfield  {author} {\bibinfo {author} {\bibfnamefont {A.}~\bibnamefont
  {Albrecht}}\ and\ \bibinfo {author} {\bibfnamefont {P.~J.}\ \bibnamefont
  {Steinhardt}},\ }\href {\doibase 10.1103/PhysRevLett.48.1220} {\bibfield
  {journal} {\bibinfo  {journal} {Phys.Rev.Lett.}\ }\textbf {\bibinfo {volume}
  {48}},\ \bibinfo {pages} {1220} (\bibinfo {year} {1982})}\BibitemShut
  {NoStop}%
\bibitem [{\citenamefont {Ade}\ \emph {et~al.}(2014)\citenamefont {Ade} \emph
  {et~al.}}]{Planck:2013jfk}%
  \BibitemOpen
  \bibfield  {author} {\bibinfo {author} {\bibfnamefont {P.~A.~R.}\
  \bibnamefont {Ade}} \emph {et~al.} (\bibinfo {collaboration} {Planck}),\
  }\href {\doibase 10.1051/0004-6361/201321569} {\bibfield  {journal} {\bibinfo
   {journal} {Astron. Astrophys.}\ }\textbf {\bibinfo {volume} {571}},\
  \bibinfo {pages} {A22} (\bibinfo {year} {2014})},\ \Eprint
  {http://arxiv.org/abs/1303.5082} {arXiv:1303.5082 [astro-ph.CO]} \BibitemShut
  {NoStop}%
\bibitem [{\citenamefont {Baumann}\ and\ \citenamefont
  {McAllister}(2015)}]{Baumann:2014nda}%
  \BibitemOpen
  \bibfield  {author} {\bibinfo {author} {\bibfnamefont {D.}~\bibnamefont
  {Baumann}}\ and\ \bibinfo {author} {\bibfnamefont {L.}~\bibnamefont
  {McAllister}},\ }\href
  {http://inspirehep.net/record/1289899/files/arXiv:1404.2601.pdf} {\emph
  {\bibinfo {title} {{Inflation and String Theory}}}}\ (\bibinfo  {publisher}
  {Cambridge University Press},\ \bibinfo {year} {2015})\ \Eprint
  {http://arxiv.org/abs/1404.2601} {arXiv:1404.2601 [hep-th]} \BibitemShut
  {NoStop}%
\bibitem [{Note1()}]{Note1}%
  \BibitemOpen
  \bibinfo {note} {There have been attempts at finding simple predictions in
  models with a few scalar fields, by identifying single field attractor
  solutions \cite {Frazer:2013zoa, Kaiser:2013sna, Kallosh:2013daa}. In the
  context of N-flation, it has also been shown that predictions can simplify in
  the large $N_{f}$ regime~\cite {Easther:2013rva,Bachlechner:2014hsa}. In
  contrast to the setup studied in this paper however, the N-flation potential
  has an exceedingly simple functional form.}\BibitemShut {Stop}%
\bibitem [{\citenamefont {Marsh}\ \emph {et~al.}(2013)\citenamefont {Marsh},
  \citenamefont {McAllister}, \citenamefont {Pajer},\ and\ \citenamefont
  {Wrase}}]{DBM}%
  \BibitemOpen
  \bibfield  {author} {\bibinfo {author} {\bibfnamefont {M.~C.~D.}\
  \bibnamefont {Marsh}}, \bibinfo {author} {\bibfnamefont {L.}~\bibnamefont
  {McAllister}}, \bibinfo {author} {\bibfnamefont {E.}~\bibnamefont {Pajer}}, \
  and\ \bibinfo {author} {\bibfnamefont {T.}~\bibnamefont {Wrase}},\ }\href
  {\doibase 10.1088/1475-7516/2013/11/040} {\bibfield  {journal} {\bibinfo
  {journal} {JCAP}\ }\textbf {\bibinfo {volume} {1311}},\ \bibinfo {pages}
  {040} (\bibinfo {year} {2013})},\ \Eprint {http://arxiv.org/abs/1307.3559}
  {arXiv:1307.3559 [hep-th]} \BibitemShut {NoStop}%
\bibitem [{\citenamefont {Frazer}\ and\ \citenamefont
  {Liddle}(2012)}]{Frazer:2011br}%
  \BibitemOpen
  \bibfield  {author} {\bibinfo {author} {\bibfnamefont {J.}~\bibnamefont
  {Frazer}}\ and\ \bibinfo {author} {\bibfnamefont {A.~R.}\ \bibnamefont
  {Liddle}},\ }\href {\doibase 10.1088/1475-7516/2012/02/039} {\bibfield
  {journal} {\bibinfo  {journal} {JCAP}\ }\textbf {\bibinfo {volume} {1202}},\
  \bibinfo {pages} {039} (\bibinfo {year} {2012})},\ \Eprint
  {http://arxiv.org/abs/1111.6646} {arXiv:1111.6646 [astro-ph.CO]} \BibitemShut
  {NoStop}%
\bibitem [{\citenamefont {Easther}\ \emph {et~al.}(2014)\citenamefont
  {Easther}, \citenamefont {Frazer}, \citenamefont {Peiris},\ and\
  \citenamefont {Price}}]{Easther:2013rva}%
  \BibitemOpen
  \bibfield  {author} {\bibinfo {author} {\bibfnamefont {R.}~\bibnamefont
  {Easther}}, \bibinfo {author} {\bibfnamefont {J.}~\bibnamefont {Frazer}},
  \bibinfo {author} {\bibfnamefont {H.~V.}\ \bibnamefont {Peiris}}, \ and\
  \bibinfo {author} {\bibfnamefont {L.~C.}\ \bibnamefont {Price}},\ }\href
  {\doibase 10.1103/PhysRevLett.112.161302} {\bibfield  {journal} {\bibinfo
  {journal} {Phys. Rev. Lett.}\ }\textbf {\bibinfo {volume} {112}},\ \bibinfo
  {pages} {161302} (\bibinfo {year} {2014})},\ \Eprint
  {http://arxiv.org/abs/1312.4035} {arXiv:1312.4035 [astro-ph.CO]} \BibitemShut
  {NoStop}%
\bibitem [{\citenamefont {Battefeld}\ and\ \citenamefont
  {Modi}(2015)}]{Battefeld:2014qoa}%
  \BibitemOpen
  \bibfield  {author} {\bibinfo {author} {\bibfnamefont {T.}~\bibnamefont
  {Battefeld}}\ and\ \bibinfo {author} {\bibfnamefont {C.}~\bibnamefont
  {Modi}},\ }\href {\doibase 10.1088/1475-7516/2015/03/010} {\bibfield
  {journal} {\bibinfo  {journal} {JCAP}\ }\textbf {\bibinfo {volume} {1503}},\
  \bibinfo {pages} {010} (\bibinfo {year} {2015})},\ \Eprint
  {http://arxiv.org/abs/1409.5135} {arXiv:1409.5135 [hep-th]} \BibitemShut
  {NoStop}%
\bibitem [{\citenamefont {{Dyson}}(1962)}]{1962JMP.....3.1191D}%
  \BibitemOpen
  \bibfield  {author} {\bibinfo {author} {\bibfnamefont {F.~J.}\ \bibnamefont
  {{Dyson}}},\ }\href {\doibase 10.1063/1.1703862} {\bibfield  {journal}
  {\bibinfo  {journal} {J. Math. Phys.}\ }\textbf {\bibinfo {volume} {3}},\
  \bibinfo {pages} {1191} (\bibinfo {year} {1962})}\BibitemShut {NoStop}%
\bibitem [{\citenamefont {Uhlenbeck}\ and\ \citenamefont
  {Ornstein}(1930)}]{Uhlenbeck:1930zz}%
  \BibitemOpen
  \bibfield  {author} {\bibinfo {author} {\bibfnamefont {G.}~\bibnamefont
  {Uhlenbeck}}\ and\ \bibinfo {author} {\bibfnamefont {L.}~\bibnamefont
  {Ornstein}},\ }\href {\doibase 10.1103/PhysRev.36.823} {\bibfield  {journal}
  {\bibinfo  {journal} {Phys.Rev.}\ }\textbf {\bibinfo {volume} {36}},\
  \bibinfo {pages} {823} (\bibinfo {year} {1930})}\BibitemShut {NoStop}%
\bibitem [{Note2()}]{Note2}%
  \BibitemOpen
  \bibinfo {note} {These can be the equations of motion either for field-field,
  field-momenta and momenta-momenta correlations function, as in \cite {2pf,
  optics, mulryne}, or for the two-index mode functions \cite {Salopek,
  Ringeval, Huston, modecode}.}\BibitemShut {Stop}%
\bibitem [{Note3()}]{Note3}%
  \BibitemOpen
  \bibinfo {note} {We will present the derivation in a separate paper \cite
  {FutureWork} but a similar calculation can be found in Refs.~\cite
  {vernizzi,battelfeld}.}\BibitemShut {Stop}%
\bibitem [{\citenamefont {Seery}\ \emph {et~al.}(2012)\citenamefont {Seery},
  \citenamefont {Mulryne}, \citenamefont {Frazer},\ and\ \citenamefont
  {Ribeiro}}]{optics}%
  \BibitemOpen
  \bibfield  {author} {\bibinfo {author} {\bibfnamefont {D.}~\bibnamefont
  {Seery}}, \bibinfo {author} {\bibfnamefont {D.~J.}\ \bibnamefont {Mulryne}},
  \bibinfo {author} {\bibfnamefont {J.}~\bibnamefont {Frazer}}, \ and\ \bibinfo
  {author} {\bibfnamefont {R.~H.}\ \bibnamefont {Ribeiro}},\ }\href {\doibase
  10.1088/1475-7516/2012/09/010} {\bibfield  {journal} {\bibinfo  {journal}
  {JCAP}\ }\textbf {\bibinfo {volume} {1209}},\ \bibinfo {pages} {010}
  (\bibinfo {year} {2012})},\ \Eprint {http://arxiv.org/abs/1203.2635}
  {arXiv:1203.2635 [astro-ph.CO]} \BibitemShut {NoStop}%
\bibitem [{\citenamefont {Dias}\ \emph {et~al.}()\citenamefont {Dias},
  \citenamefont {Frazer},\ and\ \citenamefont {Marsh}}]{FutureWork}%
  \BibitemOpen
  \bibfield  {author} {\bibinfo {author} {\bibfnamefont {M.}~\bibnamefont
  {Dias}}, \bibinfo {author} {\bibfnamefont {J.}~\bibnamefont {Frazer}}, \ and\
  \bibinfo {author} {\bibfnamefont {M.~D.}\ \bibnamefont {Marsh}},\ }\href@noop
  {} {}\bibinfo {note} {{\emph{to appear}}}\BibitemShut {NoStop}%
\bibitem [{\citenamefont {Dias}\ and\ \citenamefont
  {Seery}(2012)}]{transportns}%
  \BibitemOpen
  \bibfield  {author} {\bibinfo {author} {\bibfnamefont {M.}~\bibnamefont
  {Dias}}\ and\ \bibinfo {author} {\bibfnamefont {D.}~\bibnamefont {Seery}},\
  }\href {\doibase 10.1103/PhysRevD.85.043519} {\bibfield  {journal} {\bibinfo
  {journal} {Phys.Rev.}\ }\textbf {\bibinfo {volume} {D85}},\ \bibinfo {pages}
  {043519} (\bibinfo {year} {2012})},\ \Eprint {http://arxiv.org/abs/1111.6544}
  {arXiv:1111.6544 [astro-ph.CO]} \BibitemShut {NoStop}%
\bibitem [{\citenamefont {Dias}\ \emph {et~al.}(2012)\citenamefont {Dias},
  \citenamefont {Frazer},\ and\ \citenamefont {Liddle}}]{dbrane}%
  \BibitemOpen
  \bibfield  {author} {\bibinfo {author} {\bibfnamefont {M.}~\bibnamefont
  {Dias}}, \bibinfo {author} {\bibfnamefont {J.}~\bibnamefont {Frazer}}, \ and\
  \bibinfo {author} {\bibfnamefont {A.~R.}\ \bibnamefont {Liddle}},\
  }\href@noop {} {\  (\bibinfo {year} {2012})},\ \Eprint
  {http://arxiv.org/abs/1203.3792} {arXiv:1203.3792 [astro-ph.CO]} \BibitemShut
  {NoStop}%
\bibitem [{\citenamefont {Dean}\ and\ \citenamefont
  {Majumdar}(2006)}]{Dean:2006wk}%
  \BibitemOpen
  \bibfield  {author} {\bibinfo {author} {\bibfnamefont {D.~S.}\ \bibnamefont
  {Dean}}\ and\ \bibinfo {author} {\bibfnamefont {S.~N.}\ \bibnamefont
  {Majumdar}},\ }\href {\doibase 10.1103/PhysRevLett.97.160201} {\bibfield
  {journal} {\bibinfo  {journal} {Phys. Rev. Lett.}\ }\textbf {\bibinfo
  {volume} {97}},\ \bibinfo {pages} {160201} (\bibinfo {year} {2006})},\
  \Eprint {http://arxiv.org/abs/cond-mat/0609651} {arXiv:cond-mat/0609651
  [cond-mat]} \BibitemShut {NoStop}%
\bibitem [{Note4()}]{Note4}%
  \BibitemOpen
  \bibinfo {note} {Note that the slow-roll equations of motion are independent
  of $ \Lambda _{\protect \rm v}$, thus $ \Lambda _{\protect \rm v}$ can be
  adjusted to normalise the power spectrum.}\BibitemShut {Stop}%
\bibitem [{Note5()}]{Note5}%
  \BibitemOpen
  \bibinfo {note} {Here and throughout we are concerned with whether or not
  $\protect \qopname \relax o{log}P_{\zeta }(\protect \qopname \relax o{log}k)$
  is linear.}\BibitemShut {Stop}%
\bibitem [{\citenamefont {Lyth}(1997)}]{Lyth:1996im}%
  \BibitemOpen
  \bibfield  {author} {\bibinfo {author} {\bibfnamefont {D.~H.}\ \bibnamefont
  {Lyth}},\ }\href {\doibase 10.1103/PhysRevLett.78.1861} {\bibfield  {journal}
  {\bibinfo  {journal} {Phys. Rev. Lett.}\ }\textbf {\bibinfo {volume} {78}},\
  \bibinfo {pages} {1861} (\bibinfo {year} {1997})},\ \Eprint
  {http://arxiv.org/abs/hep-ph/9606387} {arXiv:hep-ph/9606387 [hep-ph]}
  \BibitemShut {NoStop}%
\bibitem [{\citenamefont {Frazer}(2014)}]{Frazer:2013zoa}%
  \BibitemOpen
  \bibfield  {author} {\bibinfo {author} {\bibfnamefont {J.}~\bibnamefont
  {Frazer}},\ }\href {\doibase 10.1088/1475-7516/2014/01/028} {\bibfield
  {journal} {\bibinfo  {journal} {JCAP}\ }\textbf {\bibinfo {volume} {1401}},\
  \bibinfo {pages} {028} (\bibinfo {year} {2014})},\ \Eprint
  {http://arxiv.org/abs/1303.3611} {arXiv:1303.3611 [astro-ph.CO]} \BibitemShut
  {NoStop}%
\bibitem [{\citenamefont {Kaiser}\ and\ \citenamefont
  {Sfakianakis}(2014)}]{Kaiser:2013sna}%
  \BibitemOpen
  \bibfield  {author} {\bibinfo {author} {\bibfnamefont {D.~I.}\ \bibnamefont
  {Kaiser}}\ and\ \bibinfo {author} {\bibfnamefont {E.~I.}\ \bibnamefont
  {Sfakianakis}},\ }\href {\doibase 10.1103/PhysRevLett.112.011302} {\bibfield
  {journal} {\bibinfo  {journal} {Phys. Rev. Lett.}\ }\textbf {\bibinfo
  {volume} {112}},\ \bibinfo {pages} {011302} (\bibinfo {year} {2014})},\
  \Eprint {http://arxiv.org/abs/1304.0363} {arXiv:1304.0363 [astro-ph.CO]}
  \BibitemShut {NoStop}%
\bibitem [{\citenamefont {Kallosh}\ and\ \citenamefont
  {Linde}(2013)}]{Kallosh:2013daa}%
  \BibitemOpen
  \bibfield  {author} {\bibinfo {author} {\bibfnamefont {R.}~\bibnamefont
  {Kallosh}}\ and\ \bibinfo {author} {\bibfnamefont {A.}~\bibnamefont
  {Linde}},\ }\href {\doibase 10.1088/1475-7516/2013/12/006} {\bibfield
  {journal} {\bibinfo  {journal} {JCAP}\ }\textbf {\bibinfo {volume} {1312}},\
  \bibinfo {pages} {006} (\bibinfo {year} {2013})},\ \Eprint
  {http://arxiv.org/abs/1309.2015} {arXiv:1309.2015 [hep-th]} \BibitemShut
  {NoStop}%
\bibitem [{\citenamefont {Bachlechner}\ \emph {et~al.}(2015)\citenamefont
  {Bachlechner}, \citenamefont {Dias}, \citenamefont {Frazer},\ and\
  \citenamefont {McAllister}}]{Bachlechner:2014hsa}%
  \BibitemOpen
  \bibfield  {author} {\bibinfo {author} {\bibfnamefont {T.~C.}\ \bibnamefont
  {Bachlechner}}, \bibinfo {author} {\bibfnamefont {M.}~\bibnamefont {Dias}},
  \bibinfo {author} {\bibfnamefont {J.}~\bibnamefont {Frazer}}, \ and\ \bibinfo
  {author} {\bibfnamefont {L.}~\bibnamefont {McAllister}},\ }\href {\doibase
  10.1103/PhysRevD.91.023520} {\bibfield  {journal} {\bibinfo  {journal} {Phys.
  Rev.}\ }\textbf {\bibinfo {volume} {D91}},\ \bibinfo {pages} {023520}
  (\bibinfo {year} {2015})},\ \Eprint {http://arxiv.org/abs/1404.7496}
  {arXiv:1404.7496 [hep-th]} \BibitemShut {NoStop}%
\bibitem [{\citenamefont {Dias}\ \emph {et~al.}(2015)\citenamefont {Dias},
  \citenamefont {Frazer},\ and\ \citenamefont {Seery}}]{2pf}%
  \BibitemOpen
  \bibfield  {author} {\bibinfo {author} {\bibfnamefont {M.}~\bibnamefont
  {Dias}}, \bibinfo {author} {\bibfnamefont {J.}~\bibnamefont {Frazer}}, \ and\
  \bibinfo {author} {\bibfnamefont {D.}~\bibnamefont {Seery}},\ }\href
  {\doibase 10.1088/1475-7516/2015/12/030} {\bibfield  {journal} {\bibinfo
  {journal} {JCAP}\ }\textbf {\bibinfo {volume} {1512}},\ \bibinfo {pages}
  {030} (\bibinfo {year} {2015})},\ \Eprint {http://arxiv.org/abs/1502.03125}
  {arXiv:1502.03125 [astro-ph.CO]} \BibitemShut {NoStop}%
\bibitem [{\citenamefont {Mulryne}(2013)}]{mulryne}%
  \BibitemOpen
  \bibfield  {author} {\bibinfo {author} {\bibfnamefont {D.~J.}\ \bibnamefont
  {Mulryne}},\ }\href {\doibase 10.1088/1475-7516/2013/09/010} {\bibfield
  {journal} {\bibinfo  {journal} {JCAP}\ }\textbf {\bibinfo {volume} {1309}},\
  \bibinfo {pages} {010} (\bibinfo {year} {2013})},\ \Eprint
  {http://arxiv.org/abs/1302.3842} {arXiv:1302.3842 [astro-ph.CO]} \BibitemShut
  {NoStop}%
\bibitem [{\citenamefont {Salopek}\ \emph {et~al.}(1989)\citenamefont
  {Salopek}, \citenamefont {Bond},\ and\ \citenamefont {Bardeen}}]{Salopek}%
  \BibitemOpen
  \bibfield  {author} {\bibinfo {author} {\bibfnamefont {D.}~\bibnamefont
  {Salopek}}, \bibinfo {author} {\bibfnamefont {J.}~\bibnamefont {Bond}}, \
  and\ \bibinfo {author} {\bibfnamefont {J.~M.}\ \bibnamefont {Bardeen}},\
  }\href {\doibase 10.1103/PhysRevD.40.1753} {\bibfield  {journal} {\bibinfo
  {journal} {Phys.Rev.}\ }\textbf {\bibinfo {volume} {D40}},\ \bibinfo {pages}
  {1753} (\bibinfo {year} {1989})}\BibitemShut {NoStop}%
\bibitem [{\citenamefont {Ringeval}(2008)}]{Ringeval}%
  \BibitemOpen
  \bibfield  {author} {\bibinfo {author} {\bibfnamefont {C.}~\bibnamefont
  {Ringeval}},\ }\href {\doibase 10.1007/978-3-540-74353-8_7} {\bibfield
  {journal} {\bibinfo  {journal} {Lect.Notes Phys.}\ }\textbf {\bibinfo
  {volume} {738}},\ \bibinfo {pages} {243} (\bibinfo {year} {2008})},\ \Eprint
  {http://arxiv.org/abs/astro-ph/0703486} {arXiv:astro-ph/0703486 [astro-ph]}
  \BibitemShut {NoStop}%
\bibitem [{\citenamefont {Huston}\ and\ \citenamefont
  {Christopherson}(2012)}]{Huston}%
  \BibitemOpen
  \bibfield  {author} {\bibinfo {author} {\bibfnamefont {I.}~\bibnamefont
  {Huston}}\ and\ \bibinfo {author} {\bibfnamefont {A.~J.}\ \bibnamefont
  {Christopherson}},\ }\href {\doibase 10.1103/PhysRevD.85.063507} {\bibfield
  {journal} {\bibinfo  {journal} {Phys.Rev.}\ }\textbf {\bibinfo {volume}
  {D85}},\ \bibinfo {pages} {063507} (\bibinfo {year} {2012})},\ \Eprint
  {http://arxiv.org/abs/1111.6919} {arXiv:1111.6919 [astro-ph.CO]} \BibitemShut
  {NoStop}%
\bibitem [{\citenamefont {Price}\ \emph {et~al.}(2015)\citenamefont {Price},
  \citenamefont {Frazer}, \citenamefont {Xu}, \citenamefont {Peiris},\ and\
  \citenamefont {Easther}}]{modecode}%
  \BibitemOpen
  \bibfield  {author} {\bibinfo {author} {\bibfnamefont {L.~C.}\ \bibnamefont
  {Price}}, \bibinfo {author} {\bibfnamefont {J.}~\bibnamefont {Frazer}},
  \bibinfo {author} {\bibfnamefont {J.}~\bibnamefont {Xu}}, \bibinfo {author}
  {\bibfnamefont {H.~V.}\ \bibnamefont {Peiris}}, \ and\ \bibinfo {author}
  {\bibfnamefont {R.}~\bibnamefont {Easther}},\ }\href {\doibase
  10.1088/1475-7516/2015/03/005} {\bibfield  {journal} {\bibinfo  {journal}
  {JCAP}\ }\textbf {\bibinfo {volume} {1503}},\ \bibinfo {pages} {005}
  (\bibinfo {year} {2015})},\ \Eprint {http://arxiv.org/abs/1410.0685}
  {arXiv:1410.0685 [astro-ph.CO]} \BibitemShut {NoStop}%
\bibitem [{\citenamefont {Vernizzi}\ and\ \citenamefont
  {Wands}(2006)}]{vernizzi}%
  \BibitemOpen
  \bibfield  {author} {\bibinfo {author} {\bibfnamefont {F.}~\bibnamefont
  {Vernizzi}}\ and\ \bibinfo {author} {\bibfnamefont {D.}~\bibnamefont
  {Wands}},\ }\href {\doibase 10.1088/1475-7516/2006/05/019} {\bibfield
  {journal} {\bibinfo  {journal} {JCAP}\ }\textbf {\bibinfo {volume} {0605}},\
  \bibinfo {pages} {019} (\bibinfo {year} {2006})},\ \Eprint
  {http://arxiv.org/abs/astro-ph/0603799} {arXiv:astro-ph/0603799 [astro-ph]}
  \BibitemShut {NoStop}%
\bibitem [{\citenamefont {Battefeld}\ and\ \citenamefont
  {Easther}(2007)}]{battelfeld}%
  \BibitemOpen
  \bibfield  {author} {\bibinfo {author} {\bibfnamefont {T.}~\bibnamefont
  {Battefeld}}\ and\ \bibinfo {author} {\bibfnamefont {R.}~\bibnamefont
  {Easther}},\ }\href {\doibase 10.1088/1475-7516/2007/03/020} {\bibfield
  {journal} {\bibinfo  {journal} {JCAP}\ }\textbf {\bibinfo {volume} {0703}},\
  \bibinfo {pages} {020} (\bibinfo {year} {2007})},\ \Eprint
  {http://arxiv.org/abs/astro-ph/0610296} {arXiv:astro-ph/0610296 [astro-ph]}
  \BibitemShut {NoStop}%
\end{thebibliography}%

\end{document}